\documentclass{amsart}
\usepackage{amsfonts,amsmath,amssymb,amsthm,mathtools}
\usepackage[noadjust]{cite}

\usepackage{enumerate}

\usepackage[textsize=tiny]{todonotes}

\usepackage{color}

\numberwithin{equation}{section}

\DeclareMathOperator{\Ran}{Ran}

\DeclareMathOperator{\spec}{\sigma}

\DeclareSymbolFont{SY}{U}{psy}{m}{n}
\DeclareMathSymbol{\emptyset}{\mathord}{SY}{'306}

\DeclarePairedDelimiter{\abs}{|}{|}
\DeclarePairedDelimiter{\norm}{\lVert}{\rVert}

\newcommand{\ee}{\mathrm e}

\newcommand{\eps}{\varepsilon}

\newcommand{\NN}{\mathbb{N}}
\newcommand{\RR}{\mathbb{R}}

\newcommand{\ZZ}{\mathbb{Z}}

\newcommand{\PP}{\mathbb{P}}
\newcommand{\EE}{\mathbb{E}}

\newcommand{\per}{{\mathrm{per}}}

\newcommand{\erg}{{\mathrm{erg}}}

\newcommand{\loc}{{\mathrm{loc}}}
\newcommand{\odd}{{\mathrm{odd}}}

\newcommand{\pp}{{\mathrm{pp}}}

\newtheorem{theorem}{Theorem}[section]{\bf}{\it}
{\bf}{\it}
{\bf}{\it}
\newtheorem{lemma}[theorem]{Lemma}{\bf}{\it}
\newtheorem{corollary}[theorem]{Corollary}{\bf}{\it}
\newtheorem{proposition}[theorem]{Proposition}{\bf}{\it}
{\bf}{\it}
\newtheorem{example}[theorem]{Example}{\bf}{\it}

\theoremstyle{definition}
\newtheorem{remark}[theorem]{Remark}

\title{Band edge localization beyond regular Floquet eigenvalues}
\subjclass[2010]{Primary 47B80; Secondary 35J10, 35R60, 81Q15, 82B44}
\keywords{Anderson localization, unique continuation, initial scale estimate, regular Floquet eigenvalues, Lifshitz tails}
\date{}

\author[A.~Seelmann]{Albrecht Seelmann}
\address{A.~Seelmann,
Fakult\"at f\"ur Mathematik, Technische Univer\-si\-t\"at Dortmund,
D-44221 Dortmund, Germany}
\email{albrecht.seelmann@mathematik.tu-dortmund.de}

\author[M. T\"aufer,]{Matthias T\"aufer}
\address{M. T\"aufer,
School of Mathematical Sciences,
Queen Mary University of London,
London, UK}
\email{m.taeufer@qmul.ac.uk}

\begin{document}

\begin{abstract}
 We prove that localization near band edges of multi-dimensional ergodic random Schr\"o\-dinger operators with periodic background
 potential in $L^2(\mathbb{R}^d)$ is universal. By this we mean that localization in its strongest dynamical form holds without
 extra assumptions on the random variables and independently of regularity or degeneracy of the Floquet eigenvalues of the
 background operator. The main novelty is an initial scale estimate the proof of which avoids Floquet theory altogether and uses
 instead an interplay between quantitative unique continuation and large deviation estimates. Furthermore, our reasoning is
 sufficiently flexible to prove this initial scale estimate in a non-ergodic setting, which promises to be an ingredient for
 understanding band edge localization also in these situations.
\end{abstract}

\maketitle

\section{Introduction and results}
\label{sec:intro}
The Anderson model dates back to the work of Anderson in 1958~\cite{Anderson-58} in condensed matter physics who argued that the
presence of disorder will drastically change the dynamics of electrons in a solid. 
This has triggered a huge research activity in mathematics and physics during the past 60 years. 
We refer to the monographs~\cite{PasturF-92,Stollmann-01,Veselic-08,AizenmanW-16} for an overview on the mathematics literature.
While Anderson's original work was on a lattice model, analogous phenomena have since been studied for continuum Schr\"odinger
operators. The prototypical model investigated in this context is the ergodic~\emph{Alloy-type} or~\emph{continuum Anderson model}
\[
 H_\omega^\erg = - \Delta + V_\per + V_\omega^\erg = H_\per + V_\omega^\erg,
 \quad
 V_\omega^\erg
 =
 \sum_{j \in \ZZ^d} \omega_j u(\cdot - j),
\]
in $L^2(\RR^d)$, where $V_\per$ is a $\ZZ^d$-periodic potential, $\omega=(\omega_j)_{j \in \ZZ^d}$ is a sequence of independent and
identically distributed random variables with bounded density, and $u$ is a bump function modelling the effective potential around
a single atom.
Under mild assumptions, this random family of self-adjoint operators has~\emph{almost sure spectrum}, which means that there exists
a set $\Sigma \subset \RR$ such that for almost every realization of $\omega$ the random operator $H_\omega^\erg$ has spectrum
$\Sigma$.

The general philosophy is that randomness leads to~\emph{Anderson localization}, at least in a neighbourhood of the edges of
$\Sigma$, or -- in dimension one -- on the whole of $\Sigma$.
Anderson (or~\emph{exponential}) localization in an interval $I \subset \Sigma$ means that $I$ almost surely consists
of pure point spectrum of $H_\omega^\erg$, that is,
\begin{equation}
 \label{eq:AndersonLoc}
 I \subset \spec_\pp(H_\omega^\erg),\quad
 I \cap \spec_c(H_\omega^\erg) = \emptyset,
\end{equation}
and all eigenfunctions corresponding to eigenvalues in $I$ are exponentially decaying. This is a dramatic difference to the
background operator $H_\per$ which has only absolutely continuous spectrum and no eigenvalues.
There also exist stronger notions of localization such as~\emph{dynamical localization}, describing the
non-spreading of wave packets, see, e.g.~\cite{Klein-08} for an overview.
Its formally strongest form in~\cite[Definition~3.1\,(vii)]{Klein-08} (cf.~\cite[Eq.~(1.6)]{GerminetK-01}) is formulated in terms
of a sub-exponential kernel decay in Hilbert-Schmidt norm,
\begin{equation}
  \label{eq:kernelDecay}
  \EE
  \Biggl[
  \sup_{\norm{ f }_\infty \leq 1}
  \norm{ \chi_{\Lambda_1(x)} \chi_I(H_\omega^\erg) f(H_\omega^\erg)  \chi_{\Lambda_1(y)} }_{\mathrm{HS}}^2
  \Biggr]
  \le
  C_\zeta \ee^{-\abs{x-y}^\zeta}
\end{equation}
for all $x,y \in \ZZ^d$ and $\zeta \in (0,1)$. Here, the supremum is taken over Borel functions on $\RR$,
$\norm{ \,\cdot\, }_{\mathrm{HS}}$ denotes the Hilbert-Schmidt norm, $\chi_\cdot$ denotes the indicator function of a set, and
$C_\zeta>0$ is a constant depending on $\zeta$ and various model parameters.
This indeed implies Anderson localization in $I$ by the RAGE theorem, see~\cite[Section~2.5]{AizenmanENSS-06}, or, alternatively,
via~\cite[Theorem~4.2]{GerminetK-04} in combination with~\cite[Theorem~3.11]{GerminetK-01} or~\cite[Theorem~6.4]{Klein-08}. It
also yields~\emph{strong full Hilbert-Schmidt-dynamical localization} as in~\cite[Definition~3.1\,(vi)]{Klein-08}
(cf.~\cite[Eq.~(1.7)]{GerminetK-01}),
\begin{equation}
  \label{eq:fullHSdynloc}
  \EE
  \Biggl[
  \sup_{\norm{ f }_\infty \leq 1}
  \norm{ \abs{X}^{q/2} \chi_I(H_\omega^\erg) f(H_\omega^\erg)  \chi_{\Gamma} }_{\mathrm{HS}}^2
  \Biggr]
  <
  \infty
\end{equation}
for all $q>0$ and all bounded Borel sets $\Gamma \subset \RR^d$, where $X$ denotes the multiplication by $x$; see,
e.g.,~\cite[Remark~3.3]{Klein-08}. In the current setting of random Schr\"odinger operators,~\eqref{eq:kernelDecay} is in fact
equivalent to~\eqref{eq:fullHSdynloc} by~\cite[Theorem~4.2]{GerminetK-04}.
The standard method to prove~\eqref{eq:kernelDecay} is the so-called~\emph{bootstrap multi-scale analysis} from~\cite{GerminetK-01}.

The edge of $\Sigma$ where localization is most tangible is the~\emph{bottom of the spectrum}. More challenging is the situation
where the almost sure spectrum $\Sigma$ has a~\emph{band structure}. The spectrum of operators of the form $- \Delta + V_\per$
typically has a band structure, as can be seen by Floquet theory, see e.g.~\cite{Kuchment-16} for an overview.
When an ergodic random potential $V_\omega$ is added, the almost sure spectrum $\Sigma$ of $H_\omega^\erg$ will inherit this band
structure, tacitly acknowledging that $V_\omega$ is not too large such that not all spectral gaps close. It is near these edges of
$\Sigma$ where we prove localization.

In dimension $d = 1$, randomness will immediately lead to~\emph{full localization} on the whole spectrum~\cite{GoldsheidMP-77}.
In dimensions two and larger, localization in a neighbourhood of the~\emph{bottom of the spectrum} has been proved in different
models~\cite{HoldenM-84,CombesH-94,KirschSS-98a,KloppLNS-12,GerminetHK-07,GerminetMRM-15,Klein-13,NakicTTV-18,RojasMolina-Thesis}.
Apart from the approach in~\cite{Klein-13}, which relies on sufficiently high disorder, and the one
in~\cite[Section 4.5]{RojasMolina-Thesis} for Bernoulli-Delone-operators, which relies on similar techniques to the ones we employ
below, a common strategy in this context is the utilization of so-called~\emph{Lifshitz tails} at the bottom of the spectrum. These
imply an~\emph{initial scale estimate} (ISE), a major ingredient for the above mentioned multi-scale analysis.
At internal band edges, the validity of Lifshitz tails on a general scope is a complicated issue.

Localization at internal band edges has been intensively studied in the second half of the 90s of the last century. First results,
among them~\cite{BarbarouxCH-97} by Barbaroux, Combes, and Hislop and~\cite{KirschSS-98a} by Kirsch, Stollmann, and Stolz,
additionally required a sufficiently fast decay of the distribution of the random variables $\omega_j$ near their extreme values,
for which however no physical justification is given and which excludes for instance the uniform distribution. It rather is a technical assumption necessary for the proposed proofs for an
initial scale estimate which proceed along the following idea: Consider the event where all coupling constants in a box are above a
certain threshold, thus lifting the eigenvalues, and then tune the probability distributions of the individual coupling constants
such that the probability of said event is a priori large.
In some regard, our approach (as well as the one in~\cite{RojasMolina-Thesis}) can be understood as a refinement of the technique
of~\cite{BarbarouxCH-97, KirschSS-98a}, see Remark~\ref{rem:compare} below.

Since then, further progress towards band edge localization has been driven by progress in Lifshitz tails.
The fundamental task of understanding Lifshitz tails~\emph{at internal band edges} was approached by Klopp~\cite{Klopp-99}. 
Lifshitz tails at $E_0$ mean that the integrated density of states $N(\cdot)$ of $H_\omega^\erg$ satisfies
\begin{equation}
 \label{eq:Lifshitz-tails}
 \lim_{E \searrow E_0} 
 \frac{\ln \lvert \ln \left( N(E) - N(E_0) \right) \rvert}{\ln \left(E - E_0 \right)}
 =
 - \frac{d}{2}.
\end{equation}
Klopp proved that Lifshitz tails occur for the random operator $H_\omega^\erg$ if the background operator $H_\per$
has~\emph{regular Floquet eigenvalues} near these edges, which means that these edges are generated by a quadratic extremum of an
eigenvalue curve in the so-called dispersion relation.
This implies an initial scale estimate which, together with a Wegner estimate, can then be used to start the multi-scale analysis
and prove localization~\cite{Veselic-02}. 
In fact, for compactly supported single-site potentials $u$, regular Floquet eigenvalues are even
equivalent to Lifshitz tails in the sense of~\eqref{eq:Lifshitz-tails}, see~\cite{Klopp-99}, whereas for non-compactly supported
$u$, this equivalence no longer holds since there exists another mechanism which leads to Lifshity tail
behaviour~\cite{Klopp-02}: For sufficiently slowly decaying $u$, the effective random potential around a
particular site will be an infinite, weighted average which has thin tails and will thus be of a similar form as assumed
in~\cite{BarbarouxCH-97, KirschSS-98a}.

There are now two natural questions: Firstly, is it actually possible that Floquet eigenvalues of $H_\per $ are not regular?
Secondly, how does the integrated density of states for $H_\omega^\erg$ look like if $H_\per$ exhibits a non-quadratic Floquet
minimum?

The first question is answered in dimension one and in dimension two for ``small'' potentials by~\cite{ColindeVerdiere-91}, where it is
proved that regularity of Floquet eigenvalues is generic (it occurs in a precise sense for almost all choices of the potential
$V_\per$, but there are exceptional cases where it does not).
In higher dimensions there are partial results, e.g.~\cite{KloppR-00} which states that potentials for which band edges are attained
by a single Floquet eigenvalue are generic, but the question whether regular Floquet eigenvalues are generic in all dimensions is
still open~\cite[Conjecture 5.25]{Kuchment-16}.

The second question was studied by Klopp and Wolff~\cite{KloppW-02}.
Therein it is shown in two space dimensions that even if a proper Lifshitz tail does not occur for the integrated density of states
of the random operator, a weaker version of~\eqref{eq:Lifshitz-tails} with $- d/2$ replaced by $- \alpha$ for some $\alpha > 0$
always holds.
Such an asymptotic still implies an initial scale estimate and thus localization, see Theorems~0.3 and~0.4 in~\cite{KloppW-02}.
While it seems plausible for the statement of~\cite{KloppW-02} to hold also in higher dimensions, the reasoning therein is explicitly two-dimensional
and relies on tools from analytic geometry such as the Newton diagram, which would at least introduce additional technical
complications in higher dimensions. We are not aware of any progress made following this strategy since the early 2000s. In summary, in higher dimensions band edge localization has been known only under additional assumptions so far.

Our main result of this note, Theorem~\ref{thm:Dynamical_localization}, proves that in dimension $d \geq 2$, band edge localization always occurs,
independently of the regularity or degeneracy of the Floquet eigenvalues and of Lifshitz tails. Recall that $d \geq 2$ is not a
restriction here since in dimension $d = 1$ one anyway has the stronger full localization.
The key idea comes from the observation that certain favourable configurations of the random potential have overwhelming
probability.
The scale-free quantitative unique continuation principle (UCP) for spectral
projectors from~\cite{NakicTTV-18} then allows to prove that these configurations lift eigenvalues. 
This results in a robust initial scale estimate the proof of which does not rely
on Floquet theory and makes no use of periodicity. 
Quantitative unique continuation is a technique which has been introduced to the random
Schr\"odinger operators community in~\cite{BourgainK-05} and has since found various applications in the theory of random
Schr\"odinger operators~\cite{BourgainK-13,RojasMolinaV-13,Klein-13,NakicTTV-18,TaeuferT-18,Taeufer-PhD,DietleinGM-19,Gebert-19,
RojasMolinaM-20}.

Freed from the burdens of periodicity and ergodicity, we can even state our initial scale estimate in a more general context of
non-ergodic Schr\"odinger operators in Theorem~\ref{thm:ISE}. There has recently been some activity on localization for non-ergodic
operators~\cite{RojasMolina-12,RojasMolinaV-13,Klein-13,GerminetMRM-15, TaeuferT-18, RojasMolinaM-20}.
Our initial scale estimate in Theorem~\ref{thm:ISE} might be used as an induction anchor for the multi-scale analysis for
such operators, but one would have to combine this with corresponding considerations on the multi-scale analysis in the non-ergodic
setting itself and on almost sure statements on the spectrum. This is a subject for future investigations.
Especially the existence of almost sure band edges for such operators is a somewhat touchy business.
In our context this is bypassed by Hypothesis~(H3') below, see also Remark~\ref{rem:after_hypotheses}.
At the end of Section~\ref{sec:ISE_non-ergodic}, we finally point out some situations of non-ergodic random Schr\"odinger operators
where our initial scale estimate can be used as in input.%

The paper is organized as follows: 
In Section~\ref{subsec:model}, the notation and the ergodic model are introduced whereas Section~\ref{subsec:result_ergodic}
presents the main result, Theorem~\ref{thm:Dynamical_localization}, on band edge localization.
After that, Section~\ref{sec:ISE_non-ergodic} presents Theorem~\ref{thm:ISE}, the initial scale estimate in the non-ergodic setting.
Finally, Section~\ref{sec:proof} is devoted to the proof of Theorem~\ref{thm:ISE}.

\subsection{The model}
\label{subsec:model}

We always work in dimension $d \geq 2$.
For $L > 0$ and $x \in \RR^d$, we denote by $\Lambda_L(x)$ the open hypercube in $\RR^d$ of side length $L$, centered at $x$. 
If $x = 0$, we simply write $\Lambda_L$.
Similarly, we denote by $B_\delta(x)$ the open ball of radius $\delta$, centered at $x$, and if $x = 0$ we just write $B_\delta$.
Given a measurable subset $A \subset \RR^d$ we write $\chi_A$ for the characteristic function of this set. 

We consider a $\ZZ^d$-ergodic random Schr\"odinger operator $H_\omega^\erg = H_\per + V_\omega^\erg$ in $L^2(\RR^d)$
satisfying the following hypotheses (cf., e.g.,~\cite{KirschSS-98a,DamanikS-01}):

\begin{enumerate}[({H}1)]
 \item
 $H_\per = - \Delta + V_\per$, where $V_\per \in L^\infty(\RR^d)$ is $\ZZ^d$-periodic and real-valued.
 \item
 $V_\omega^\erg = \sum_{j \in \ZZ^d} \omega_j u(\cdot - j)$, where $u \in L^p(\RR^d)$ with $p = 2$ if $d \in \{2,3\}$ and $p > d/2$
 if $d \geq 4$ is nonnegative and compactly supported. Moreover, there are $c, \delta > 0$ such that
 \[
  c \chi_{B_\delta} \leq u.
 \]
 The random variables $\omega_j$ are
 independent and identically distributed on a probability space $(\Omega, \PP)$ with bounded density and support equal to the interval $[0,1]$.
 \item
 There are $- \infty \leq a < b < \infty$ such that $(a,b) \in \rho(H_\omega^\erg)$ and $b \in \sigma(H_\omega^\erg)$ almost surely.
\end{enumerate}

The reason why we assume the potential $V_\per$ in~(H1) to be bounded is that this is a requirement of the quantitative unique
continuation principle for spectral projectors~\cite{NakicTTV-18}, a major ingredient in the proof. There have been recent works
removing this boundedness assumption~\cite{KleinT-16,KleinT-16a}, but since this is not the main focus of this work, we refrain from
pursuing this path further here.

\subsection{Main results}
\label{subsec:result_ergodic}

The following theorem spells out localization in a neighbourhood of the lower edge of a connected component of the almost sure spectrum.
The case of an upper edge can be treated analogously by obvious modifications to Hypothesis (H3) and the proofs.

\begin{theorem}[Dynamical localization near band edges]
  \label{thm:Dynamical_localization}
  Assume (H1), (H2), and (H3).
  Then there exists $\epsilon > 0$ such that for all $\zeta \in (0,1)$ there is a constant $C_\zeta > 0$ with
  \begin{equation*}
    \EE
    \Biggl[
    \sup_{\norm{ f }_\infty \leq 1}
    \norm{ \chi_{\Lambda_1(x)} \chi_{[b, b + \epsilon]}(H_\omega) f(H_\omega)  \chi_{\Lambda_1(y)} }_{\mathrm{HS}}^2
    \Biggr]
    \le
    C_\zeta \ee^{-\abs{x-y}^\zeta}
  \end{equation*}
  for all $x,y \in \ZZ^d$. Here, the supremum is taken over all Borel functions on $\RR$, and $\norm{ \,\cdot\, }_{\mathrm{HS}}$
  denotes the Hilbert-Schmidt norm.
\end{theorem}

As mentioned in the introduction, Theorem~\ref{thm:Dynamical_localization} yields strong Hilbert-Schmidt-dynamical localization,
as well as Anderson localization, in the interval $[b,b+\eps]$, cf.~\eqref{eq:fullHSdynloc} and~\eqref{eq:AndersonLoc},
respectively. It also includes the essentially well-known particular case where the bottom of the almost sure spectrum is
considered, cf.~\cite{KirschSS-98a,DamanikS-01}.
On the other hand, to the best of our knowledge,~Theorem~\ref{thm:Dynamical_localization} provides the first proof of such
localization near internal band edges in the continuum without additional assumptions. In particular, it does not require
regularity of Floquet eigenvalues as in~\cite{Veselic-02}.
The core of the proof of Theorem~\ref{thm:Dynamical_localization} is a so-called~\emph{initial scale estimate}, which in the
situation of the theorem takes the following form (see also Corollary~\ref{cor:ergISE} below):

For all $q > 0$ and $\alpha \in (0,1)$ there exists $L_0 \in \NN$ such that for all $L \in \NN$ with $L \geq L_0$ we have
\begin{equation}
  \label{eq:ergISE}
  \PP
  \bigl[
  \sigma ( H_{\omega,L}^\erg) \cap [b, b + L^{- \alpha}) = \emptyset 
  \bigr]
  \geq 
  1 - L^{- q},
\end{equation}
where $H_{\omega,L}^\erg$ denotes the restriction of $H_\omega^\erg$ onto $L^2(\Lambda_L)$ with periodic boundary conditions.
Theorem~\ref{thm:Dynamical_localization} then follows from~\eqref{eq:ergISE}, combined with the well-known Combes-Thomas estimate,
and the Wegner estimate~\cite[Theorem~3.1]{KirschSS-98a} via the bootstrap multi-scale analysis, see,
e.g.,~\cite[Theorem~3.8]{GerminetK-01}; cf.~also~\cite[Theorems~4.1 and~4.2]{GerminetK-04}. Since this procedure is well understood
by now, we omit the explicit treatment of this multi-scale analysis here and just content ourselves with the proof of the initial
scale estimate~\eqref{eq:ergISE}. 

\begin{remark}
 Our proof of the initial scale estimate does not rely on the fact that the $\omega_j$ have a  Lebesgue density. 
This opens the way to consider band edge localization for, e.g., i.i.d.~Bernoulli random variables. 
 In this situation, a Wegner estimate is no longer available, but an initial scale estimate still implies localization via the
 multi-scale analysis developed in~\cite{BourgainK-05}, see for instance Remark~1.3 of~\cite{GerminetK-13} for an explicit
 reference in this setting with bounded single-site potential $u$.
\end{remark}

Although it is very plausible that~\emph{dynamical localization} should hold also in the case where the single-site potential is
not compactly supported but has sufficiently fast decay at infinity, we are not aware of any proofs for the corresponding
multi-scale analysis in the literature; we note however that a partial result has been stated in~\cite[Theorem~4.3]{DamanikS-01}.
On the other hand, for only \emph{Anderson localization} the multi-scale analysis in this long range setting is well established,
see, e.g.,~\cite{KirschSS-98b,Zenk-02,LeonhardtPTV-15}. Since the proof of our initial scale estimate is not restricted to
compactly supported single-site potentials, we therefore still have the following statement:
\begin{remark}
 If the single-site potential $u$ in~(H2) is not assumed to have compact support but merely to have sufficiently fast decay at
 infinity, the operator $H_\omega^\erg$ still exhibits Anderson localization near band edges, cf.~\cite{KirschSS-98b,Veselic-02}.
\end{remark}

Our proof of the initial scale estimate does also not rely on periodicity or ergodicity. Therefore, we prove a more general initial
scale estimate for not necessarily ergodic random Schr\"odinger operators $H_\omega = H_0 + V_\omega$ in Theorem~\ref{thm:ISE}
below.

Let us conclude this section by briefly explaining the main idea of the proof of the initial scale estimate~\eqref{eq:ergISE}; the full proof in the more general non-ergodic situation can be found in Section~\ref{sec:proof} below:
The quantitative unique continuation principle implies that eigenvalues of $- \Delta + V$ will move up by some positive amount if the potential $V$ is varied by some $c > 0$ on a -- in general disconnected -- subset $U$ with typical distance $l$ between its components.
However, the price to pay is that the lifting is very small and it scales unfavourably with increasing $l$, namely the eigenvalues will be lifted proportional to $c \exp(- l^{4/3+\eps})$.
On the one hand, this seems to be too weak for the polynomial bound in \eqref{eq:ergISE}.
On the other hand, by large deviation arguments favourable configurations of the random potential appear with overwhelming probability for large $l$: The random potential will be larger than $c$ on a set $U$ with typical distance $l$ between its components with probability $1 - \exp(- l^d)$.
Since $\exp(- l^{4/3+\eps})$ decays slower than $\exp(- l^d)$ in dimensions $d \geq 2$, we can trade the large deviations bound against the meager eigenvalue lifting from unique continuation and conclude the statement.

\section{An initial scale estimate for non-ergodic random Schr\"odinger operators}
\label{sec:ISE_non-ergodic}

 For a random operator $H_\omega = H_0 + V_\omega$, $L > 0$, and $x \in \RR^d$, we denote by $H_{\omega, L, x}$ the restriction of $H_\omega$ to
 $L^2(\Lambda_L(x))$ with a fixed choice of either Dirichlet, Neumann, or periodic boundary conditions.

We formulate the following hypotheses:
\begin{enumerate}[({H}1')]
 \item
 $H_0 = - \Delta + V_0$, where $V_0 \in L^\infty(\RR^d)$ is real-valued.
 \item
 $V_\omega = \sum_{j \in (G\ZZ)^d} \omega_j u_j$, $G > 0$, where $u_j \in L_\loc^p(\RR^d)$ with $p > d/2$ is nonnegative satisfying
 \begin{equation}
  \label{eq:uniform_bound}
  \sup_{k\in(G\ZZ)^d} \sum_{j\in(G\ZZ)^d} \norm{u_j}_{L^p(\Lambda_G(k))} < \infty.
 \end{equation} 
 In addition, there are $c > 0$ and $\delta \in (0,G/2]$ such that for every $j \in (G \ZZ)^d$ there exists $x_j \in \Lambda_G(j)$
 with $B_\delta(x_j) \subset \Lambda_G(j)$ and
 \[
  c \chi_{B_\delta(x_j)} \leq u_j.
 \]
 Moreover, the random variables $\omega_j$ are independent on a probability space $(\Omega, \PP)$, with values contained in the
 interval $[0,1]$ almost surely, and there are $\eta, \kappa > 0$ such that $\PP[ \omega_j \geq \eta ] \geq \kappa$ for all
 $j \in (G \ZZ)^d$.
 \item
 There are $b \in \RR$ and $M_b \subset (G \NN) \times (G \ZZ)^d$ such that for all $(L,x) \in M_b$ there is $a < b$ with
 $(a,b) \subset \rho(H_{0,L,x} + t W_{L,x})$ for all $t \in [0,1]$, where
 \begin{equation*}
	W := \sum_{j \in (G \ZZ)^d} u_j.
 \end{equation*}
\end{enumerate}

\begin{remark}
 \label{rem:after_hypotheses}
 (1) If the random variables $\omega_j$ in (H2') are identically distributed and non-trivial, then one can choose $\kappa = 1/2$
 and $\eta = \operatorname{Med} (\omega_0)$, the median of $\omega_0$.
 
 (2) Condition~\eqref{eq:uniform_bound} guarantees that $W$ in (H3') is locally $L^p$ with a uniform bound on the $L^p$-norm on
 cubes of side length $G$. Since $p > d/2$ and $d \ge 2$, this ensures that $W$ (and consequently also $V_\omega$ almost surely)
 belongs to the Kato class in $\RR^d$ and is thus infinitesimally form bounded with respect to $H_0$, see,
 e.g.,~\cite[Section~1.2]{CyconFKS-89}. Note that~\eqref{eq:uniform_bound} is obviously satisfied if
 $\norm{u_j}_{L^p(\Lambda_G(k))}$ exhibits a sufficiently fast decay in $\abs{j-k}$, say
 $\norm{u_j}_{L^p(\Lambda_G(k))} \le \text{const}/(1+\abs{j-k})^{m}$ for all $k,j \in (G\ZZ)^d$ and some $m > d$.
 
 (3) Hypothesis (H3') implies that for every $(L,x) \in M_b$ the number of eigenvalues of $H_{\omega,L,x}$ below $b$ is almost
 surely constant whence the infimum of the spectrum of $H_{\omega,L,x}$ in $[b, \infty)$ can experience no ``jumps'' when the
 random potential is varied, see the proof of Lemma~\ref{lem:A_moves} below for more details.
\end{remark}

The following theorem is the core of the present paper:

\begin{theorem}[ISE for non-ergodic random Schr\"odinger operators]
 \label{thm:ISE}
  Assume Hypotheses (H1'), (H2'), and (H3').
  Then, for all $q > 0$ and $\alpha \in (0,1)$ there exists $L_0 \in G \NN$ such that for all $(L,x) \in M_b$ with $L \geq L_0$ we have
  \[
    \PP
    \bigl[
    \sigma ( H_{\omega,L,x}) \cap [b, b + L^{- \alpha}) = \emptyset
    \bigr]
    \geq 
    1 - L^{- q}.
  \]
\end{theorem}

It is clear that (H1)--(H2) are a particular case of (H1')--(H2') with $V_0=V_\per$, $G=1$, and $u_j=u(\cdot-j)$.
We also note that Hypotheses (H1')--(H2') define a generalization of the~\emph{crooked Anderson model} and
the~\emph{Delone-Anderson model} for which Wegner estimates are known~\cite{Klein-13,RojasMolinaV-13}.
Let us comment on the connection between the gap conditions (H3) and (H3'):
With the fixed choice of periodic boundary conditions, (H3) implies (H3') with $M_b = \NN \times \ZZ^d$, and the gap $(a,b)$ in
(H3) belongs to the resolvent set of each periodic box restriction in (H3').
Indeed, the function $W$ in (H3') then is $\ZZ^d$-periodic and the almost sure spectrum of $H^\erg_\omega$ agrees with
$\Sigma = \bigcup_{t \in [0,1]} \spec(H_\per + t W)$, see, e.g.,~\cite[Lemma~1.4.1]{Stollmann-01}.
It is then a consequence of Floquet theory that the gap $(a,b) \subset \RR \setminus \Sigma$ also belongs to the resolvent set of
the periodic restrictions of $H_\per + tW$ to the boxes $\Lambda_L$ of integer side length in (H3').
Note that the latter reasoning also applies with Dirichlet and Neumann boundary conditions if the $\ZZ^d$-periodic background
potential $V_\per$ additionally is also symmetric under all coordinate reflections. This can be seen by extending Neumann and
Dirichlet eigenfunctions by symmetric and antisymmetric reflections, respectively, to a box of double the side length, on which
they satisfy periodic boundary conditions.

In summary, we have seen that~\eqref{eq:ergISE} is a particular case of Theorem~\ref{thm:ISE}, so that we obtain the following corollary:

\begin{corollary}[Ergodic ISE at band edges]
 \label{cor:ergISE}
 Assume (H1), (H2), and (H3).
 Then for all $q > 0$ and $\alpha \in (0,1)$ there exists $L_0 \in \NN$ such that for all $L \in \NN$ with $L \geq L_0$ we have
 \[
  \PP
  \bigl[
  \sigma ( H_{\omega,L}^\erg) \cap [b, b + L^{- \alpha}) = \emptyset 
  \bigr]
  \geq 
  1 - L^{- q}.
 \]
\end{corollary}

The fact that Theorem~\ref{thm:ISE} does not require ergodicity leads to a robustness of localization proofs which is difficult to
achieve when relying on Floquet theory. We are not going to engage in the multi-scale analysis in the non-ergodic setting here, but
we nevertheless point out some non-ergodic situations where Theorem~\ref{thm:ISE} can be used as an input, namely:

\begin{example}
 \begin{enumerate}[(a)]
 \item
 The Delone-Anderson model
 \[
  H_\omega = - \Delta + V_\per + \sum_{x \in \mathcal{D}} \omega_j u( \cdot - x)
 \]
 where $\mathcal{D} \subset \RR^d$ is a so-called~\emph{Delone} set, see~\cite{RojasMolinaV-13, Klein-13} for definitions. One may
 even consider here to have the periodic background potential $V_\per$ be perturbed by a fast decaying potential. We note however
 that an alternative approach to fast decaying perturbations can be found in~\cite{DietleinGHKM-18}, where it is proved that
 localization is robust under such perturbations.
 
 \item
 More generally, the crooked Anderson model treated in~\cite{Klein-13, TaeuferT-18}, assuming that the random variables do not
 concentrate too much, so that there exist $\eta$ and $\kappa$ as in (H2').

 \item
 Models of the form
 \[
  H_\omega = - \Delta + V_{\per,G_1} + \sum_{x \in (G_2 \ZZ)^d} \omega_j u(\cdot - j),
 \]
 where the periodicity cell of the background potential is of a size $G_1 > 0$ incommensurate with the ergodicity length $G_2$,
 that is, the system exhibits a~\emph{quasi-periodic} structure.
 \end{enumerate}
\end{example}

In the above situations, Hypothesis (H3') may be verified, for instance, by supposing that the random potential is sufficiently
small, provided that the background operator admits a suitable spectral gap for the box restrictions $H_{0,L,x}$.

\section{Proof of Theorem~\ref{thm:ISE}}
\label{sec:proof}
By scaling, we may assume $G = 1$.
Furthermore, for notational convenience, we assume that $\NN \times \{ 0 \} \subset M_b$ and therefore only prove the statement for
$x = 0$ and sufficiently large $L$, writing $H_{\omega,L}$ and $H_{0,L}$ instead of $H_{\omega,L,x}$ and $H_{0,L,x}$.

The proof of Theorem~\ref{thm:ISE} relies on the scale-free quantitative unique continuation principle~\cite{NakicSTTV-19} given in
Proposition~\ref{prop:UCP} below.
We start by introducing some notation:
Given $l>0$ and $\delta\in(0,l/2)$, a sequence $Z=(y_j)_{j\in (l\ZZ)^d}$ in $\RR^d$ is called
\emph{$(l,\delta)$-equidistributed} if $B_\delta(y_j)\subset \Lambda_l(j)$ for all $j\in(l\ZZ)^d$. 
If $Z$ is $(l, \delta)$-equidistributed and $L > 0$, we write
\[
 S_{Z,L}
 :=
 \bigcup_{j \in (l\ZZ)^d} B_\delta(y_j) \cap \Lambda_L.
\]

\begin{proposition}[Scale-free unique continuation principle~\cite{NakicTTV-18,NakicSTTV-19}]
 \label{prop:UCP}
 Let $V \in L^\infty(\RR^d)$, $l \in \NN_\odd$, $L \in \NN$ with $l \leq L$, and $\delta\in(0,l/2)$. 
 Denote the restriction of $- \Delta + V$ to
 $\Lambda_L$ with Dirichlet, Neumann, or periodic boundary conditions by $H_L$.
 Let $Z$ be an $(l,\delta)$-equidistributed sequence. 
 Then, for every $E\in\RR$ and every $\phi\in\Ran(\chi_{(-\infty,E]}(H_L))$
 we have
 \begin{equation}
  \label{eq:result1}
  \lVert \phi \rVert_{L^2 (S_{Z,L})}^2
  \geq 
  \left(\frac{\delta}{l} \right)^{N \bigl(1 + l^{4/3} \lVert V \rVert_\infty^{2/3} + l \sqrt{\max \{ E,0 \}} \bigr)}
  \lVert \phi \rVert_{L^2 (\Lambda_L)}^2,
 \end{equation}
 where $N > 0$ is a constant that only depends on the dimension.
\end{proposition}

Note also that for fixed $\delta \in (0, 1/2]$ and sufficiently large $l$ (depending on $\delta$, $\lVert V \rVert_\infty$, $E$, and $N$ only) we can estimate
\begin{equation}
 \label{eq:expDecay}
 \begin{aligned}
  \left(\frac{\delta}{l} \right)^{N \bigl(1 + l^{4/3} \lVert V \rVert_\infty^{2/3} + l \sqrt{\max\{E,0\}} \bigr)}
  &\geq
  \exp 
  \bigl(
  - l^{7/5}
  \bigr).
 \end{aligned}
\end{equation}

In the situation of Theorem~\ref{thm:ISE}, let $J(\omega):=\{ k\in \ZZ^d \colon \omega_k \ge \eta \} \subset \ZZ^d$ for $\omega\in\Omega$ and consider for
$l \leq L$ the event
\begin{equation}
 \label{eq:defA}
 A_{l,L}
 :=
 \bigl\{ \omega \in \Omega \colon \Lambda_l(j) \cap J(\omega) \neq \emptyset\ \text{ for all }j\in(l\ZZ)^d \cap \Lambda_{2L}\bigr\}.
\end{equation}

The main idea is that if $\omega \in A_{l,L}$ then we can pick $k_j \in \Lambda_l(j) \cap J(\omega)$, where $j$ runs over $(l\ZZ)^d \cap \Lambda_{2L}$, such that the
corresponding points $x_{k_j}$ from Hypothesis (H2') are part of an
$(l,\delta)$-equidistributed sequence.
The scale-free unique continuation principle then implies the following eigenvalue lifting estimate:
\begin{lemma}
 \label{lem:A_moves}
 There is $l_0 \in \NN$, depending only on $\delta$, $b$, $d$, $c$, $\eta$, and
 $\norm{V_0}_\infty$, such that for all $L \in \NN$ and $l \in \NN_\odd$ with $L \geq l \geq l_0$
 and all $\omega \in A_{l,L}$ we have
 \[
   \inf \sigma(H_{\omega, L}) \cap [b, \infty) \geq b + \eta c \exp \bigl( - l^{7/5} \bigr).
 \]
\end{lemma}

\begin{proof}
 If $\inf \sigma(H_{\omega, L}) \cap [b, \infty) \geq b + \eta c$, there is nothing to prove. 
 So, from now on assume that
 \begin{equation}
 \label{eq:E}
  \inf \sigma(H_{\omega, L}) \cap [b, \infty) \leq b + \eta c =: E.
 \end{equation}
 Since $H_{0,L} \leq H_{0,L} + \eta c\chi_{S_{Z,L}} \leq H_{\omega,L} \leq H_{0,L} + W_L$, the minimax principle for eigenvalues implies that for every $k \in \NN$ the
 $k$-th eigenvalues, counted from the bottom of the spectrum, satisfy
\begin{equation}
 \label{eq:eigenvalue_sandwich}
 \lambda_k(H_{0,L}) 
 \leq 
 \lambda_k(H_{0,L} + \eta c\chi_{S_{Z,L}}) 
 \leq
 \lambda_k(H_{\omega,L}) 
 \leq 
 \lambda_k(H_{0,L} + W_L).
\end{equation}
Observe that $t \mapsto H_{0,L} + t W_L$ defines by~\cite[Theorem~VII.4.2]{Kato-76} a so-called~\emph{holomorphic family of operators
of type (B)} in a sufficiently small complex neighbourhood of $[0,1]$. In turn, it follows from~\cite[Remark~VII.4.22]{Kato-76} that
the eigenvalue curves $t \mapsto \lambda_k( H_{0,L} + t W_L )$ with the prescribed ordering of the eigenvalues are continuous on
$[0,1]$.
Since Hypothesis~(H3') for fixed $L$ prevents eigenvalues of the family $[0,1] \mapsto H_{0,L} + t W_L$ from entering the corresponding
interval $(a,b)$ during this variation, we conclude from Ineq.~\eqref{eq:eigenvalue_sandwich} that no eigenvalues can enter $(a,b)$
when passing from $H_{0,L} + \eta c\chi_{S_{Z,L}}$ to $H_{\omega,L}$ either. 
In particular, there exists $k_0 \in \NN$ such that $\lambda_{k_0}(H_{0,L})$, $\lambda_{k_0}(H_{0,L} + \eta c\chi_{S_{Z,L}})$, and $\lambda_{k_0}(H_{\omega,L})$ denote the lowest eigenvalue in $[b, \infty)$ of the respective operators.
Therefore, it suffices to prove that
\begin{equation}
 \label{eq:eigenvalue_moving}
 \lambda_{k_0}(H_{0,L} + \eta c\chi_{S_{Z,L}})
 \geq
 \lambda_{k_0}(H_{0,L})
 +
 \eta c \exp(- l^{7/5}).
\end{equation}
To this end, observe that by~\eqref{eq:E} and~\eqref{eq:eigenvalue_sandwich} we have
\[
 \lambda_{k_0}(H_{0,L} + \eta c\chi_{S_{Z,L}})
 \leq
 \lambda_{k_0}(H_{\omega,L})
 \leq
 E.
\]
Hence, for sufficiently large $l$, Proposition~\ref{prop:UCP} and Ineq.~\eqref{eq:expDecay} yield that
\[
 \lVert \eta c\chi_{S_{Z,L}} \phi \rVert^2 
 \geq
 \eta c \exp(- l^{7/5}) \lVert \phi \rVert^2
\]
for all $\phi \in \Ran ( \chi_{(- \infty,E]}(H_{0,L} + \eta c\chi_{S_{Z,L}}))$.
Ineq.~\eqref{eq:eigenvalue_moving} now follows from the minimax principle for eigenvalues as, for instance, in Lemma 3.5 of~\cite{NakicSTTV-19}.
\end{proof}%

We are ready to prove Theorem~\ref{thm:ISE}.

\begin{proof}[Proof of Theorem~\ref{thm:ISE}]
 For sufficiently large $L \in \NN$ (depending only on $\alpha$), we find $l \in \NN_\odd$ with $l \leq L$ such that
 \begin{equation}
 \label{eq:l}
  \frac{1}{2} \ln (L^\alpha)^{2/3} < l \le \ln (L^\alpha)^{2/3}.
 \end{equation}
 Choosing $L$ possibly larger (depending only on $\alpha, \eta, c$) we furthermore have that
 $l^{7/5} \le \ln (L^\alpha)^{\frac{14}{15}} \leq \ln(c \eta L^\alpha)$, which implies
 \[
  \eta c \exp \bigl( - l^{7/5} \bigr)
  \geq
  L^{-\alpha}.
 \]
 From Lemma~\ref{lem:A_moves} we deduce
 \begin{equation}
  \label{eq:ISEbound}
  \begin{aligned}
   &\PP\bigl[ \inf \sigma(H_{\omega, L}) \cap [b, \infty) \geq b + L^{-\alpha} \bigr]\\
   \geq
   &\PP
   \Big[
   \inf \sigma(H_{\omega, L}) \cap [b, \infty) \geq b + \eta c \exp \bigl( - l^{7/5} \bigr)
   \Bigr]
   \geq
   \PP
   [A_{l,L}].
  \end{aligned}
 \end{equation}
 It remains to give a lower bound on $\PP [A_{l,L}]$. 
 To this end, note that since $l \in \NN_\odd$, we have for each $j \in \Lambda_L \cap (l\ZZ)^d$ that
 $\#(\Lambda_l(j) \cap \ZZ^d) = l^d$.
 Thus
 \[
  \PP \bigl[ \{\omega \colon \Lambda_l(j) \cap J(\omega) = \emptyset \} \bigr]
  =
  \PP \bigl[ \{\omega \colon \omega_k < \eta\ \forall k \in \Lambda_l(j) \cap \ZZ^d \} \bigr]
  \le
  ( 1 - \kappa )^{l^d}.
 \]
 Inserting~\eqref{eq:l} yields
 \begin{align*}
  \PP \bigl[ \{\omega \colon \Lambda_l(j) \cap J(\omega) &= \emptyset \} \bigr]
  \le
  ( 1 - \kappa )^{(\frac{1}{2}\ln(L^\alpha)^{2/3})^d}\\
  &=
  \exp\Bigl( \frac{\ln(1-\kappa)}{2^d} \alpha^{2d/3} \ln(L)^{(2d-3)/3} \cdot \ln L \Bigr)\\
  &\le
  \exp \bigl( -(q + d) \ln L  - d \ln 2 \ln L \bigr) 
  \\
  & \le
  \exp \bigl( -q \ln(L) - d \ln (2 L) \bigr)
  = (2L)^{-d} L^{-q},
 \end{align*}
 provided that $L$ is so large that $-\frac{\ln(1-\kappa)}{2^d} \alpha^{2d/3} \ln(L)^{(2d-3)/3} \ge (q + d) + d \ln 2$; 
 recall here that $d \ge 2$.
 
 Finally, by a union bound we obtain
 \begin{align*}
  \PP [\Omega \setminus A_{l,L}]
  &\le
  \sum_{j \in \Lambda_{2 L} \cap (l\ZZ)^d} \PP \bigl[ \{\omega \colon \Lambda_l(j) \cap J(\omega) = \emptyset \} \bigr]\\
  &\le
  (2L)^d \cdot (2L)^{-d} \cdot L^{-q} = L^{-q},
 \end{align*}
 which, in view of~\eqref{eq:ISEbound}, proves the claim.
\end{proof}

\begin{remark}
 \label{rem:compare}
 (1)
 At the bottom of the spectrum it suffices to have a quantitative unique continuation principle for eigenfunctions (and not for
 spectral subspaces) as in~\cite{RojasMolinaV-13}. In the setting of Bernoulli random variables, such an argument appears in the
 PhD thesis of Rojas-Molina, see~\cite[Section~4.5]{RojasMolina-Thesis}, and in~\cite[Lemma~3.8]{RojasMolinaM-20}.

 (2) 
 Our approach (as well as the one in~\cite{RojasMolina-Thesis}) can in some regard be understood as a refinement of the technique
 of~\cite{BarbarouxCH-97, KirschSS-98a}: We do not consider the event where \emph{all} coupling constants exceed a value, but only
 random variables in a sufficiently rich subset. Due to large deviation arguments, the probability of such configurations has thin
 tails and this eliminates the need to tune the individual probability distributions as in~\cite{BarbarouxCH-97, KirschSS-98a}.

 (3) 
 The proof of Theorem~\ref{thm:ISE} merely relies on the fact that configurations for which the potential is larger than $\eta c$
 on an $(l, \delta)$-equidistributed set within $\Lambda_L$ has overwhelming probability. Therefore, its proof verbatim transfers
 to other models which share this feature such as the~\emph{random breather model}~\cite{TaeuferV-15,SchumacherV-17,NakicTTV-18}.

 (4)
 We use that the probability for the set $\{ x \in \Lambda_l \colon V_\omega(x) \geq c \}$ to contain a $\delta$-ball is of order $1 - \exp( - l^d)$.
 A careful look at the proof shows that it would indeed suffice for this to be of order $1 - \exp( - l^{4/3 + \epsilon})$ for some $\epsilon > 0$.
 Thus, for any non-negative random potential with this property we also obtain an initial scale estimate.
 This includes for instance models based on random fields such as Gaussian processes.
\end{remark}

\subsubsection*{Acknowledgments}
The authors thank Ivan Veseli\'c for suggesting this research direction. 
Discussions with Sasha Sodin as well as valuable comments by Martin Tautenhahn are gratefully acknowledged.
The authors are also indebted to the anonymous referees for comments that helped to improve this manuscript.
The second named author was supported by the European Research Council starting grant 639305 (SPECTRUM).

\newcommand{\etalchar}[1]{$^{#1}$}


\begin{thebibliography}{GMRM15}

\bibitem[AEN{\etalchar{+}}06]{AizenmanENSS-06}
M.~Aizenman, A.~Elgart, S.~Naboko, J.~H. Schenker, and G.~Stolz.
\newblock Moment analysis for localization in random {S}chr\"{o}dinger
  operators.
\newblock {\em Invent. Math.}, 163(2):343--413, 2006.

\bibitem[And58]{Anderson-58}
P.W. Anderson.
\newblock Absence of diffusion in certain random lattices.
\newblock {\em Phys. Rev.}, 109:1492, 1958.

\bibitem[AW15]{AizenmanW-16}
M.~Aizenman and S.~Warzel.
\newblock {\em Random operators: Disorder {E}ffects on {Q}uantum {S}pectra and
  {D}ynamics}, volume 168 of {\em Graduate Studies in Mathematics}.
\newblock American Mathematical Society, Providence, RI, 2015.

\bibitem[BCH97]{BarbarouxCH-97}
J.~M. Barbaroux, J.~M. Combes, and P.~D. Hislop.
\newblock Localization near band edges for random {S}chr\"{o}dinger operators.
\newblock {\em Helv. Phys. Acta}, 70(1-2):16--43, 1997.
\newblock Papers honouring the 60th birthday of Klaus Hepp and of Walter
  Hunziker, Part II (Z\"{u}rich, 1995).

\bibitem[BK05]{BourgainK-05}
J.~Bourgain and C.~E. Kenig.
\newblock On localization in the continuous {A}nderson-{B}ernoulli model in
  higher dimension.
\newblock {\em Invent. Math.}, 161(2):389--426, 2005.

\bibitem[BK13]{BourgainK-13}
J.~Bourgain and A.~Klein.
\newblock Bounds on the density of states for {S}chr\"{o}dinger operators.
\newblock {\em Invent. Math.}, 194(1):41--72, 2013.

\bibitem[CdV91]{ColindeVerdiere-91}
Y.~Colin~de Verdi\`ere.
\newblock Sur les singularit\'{e}s de van {H}ove g\'{e}n\'{e}riques.
\newblock In {\em Analyse globale et physique math\'{e}matique (Colloque \`a la
  m\'emoire d'Edmond Combet, Lyon, 1989)}, number~46 in M\'{e}moires de la
  Soci\'{e}t\'{e} Math\'{e}matique de France, pages 99--109. Soci\'et\'e
  math\'ematique de France, 1991.

\bibitem[CFKS87]{CyconFKS-89}
H.~L. Cycon, R.~G. Froese, W.~Kirsch, and B.~Simon.
\newblock {\em {S}chr\"{o}dinger {O}perators with {A}pplication to {Q}uantum
  {M}echanics and {G}lobal {G}eometry}.
\newblock Texts and Monographs in Physics. Springer-Verlag, Berlin, study
  edition, 1987.

\bibitem[CH94]{CombesH-94}
J.-M. Combes and P.~D. Hislop.
\newblock Localization for some continuous, random {H}amiltonians in
  {$d$}-dimensions.
\newblock {\em J. Funct. Anal.}, 124(1):149--180, 1994.

\bibitem[DGH{\etalchar{+}}18]{DietleinGHKM-18}
A.~Dietlein, M.~Gebert, P.~D. Hislop, A.~Klein, and P.~M\"{u}ller.
\newblock A bound on the averaged spectral shift function and a lower bound on
  the density of states for random {S}chr\"{o}dinger operators on
  {$\mathbb{R}^d$}.
\newblock {\em Int. Math. Res. Not. IMRN}, 2018(21):6673--6697, 2018.

\bibitem[DGM19]{DietleinGM-19}
A.~Dietlein, M.~Gebert, and P.~M\"{u}ller.
\newblock Perturbations of continuum random {S}chr\"{o}dinger operators with
  applications to {A}nderson orthogonality and the spectral shift function.
\newblock {\em J. Spectr. Theory}, 9(3):921--965, 2019.

\bibitem[DS01]{DamanikS-01}
D.~Damanik and P.~Stollmann.
\newblock Multi-scale analysis implies strong dynamical localization.
\newblock {\em Geom. Funct. Anal.}, 11(1):11--29, 2001.

\bibitem[Geb19]{Gebert-19}
M.~Gebert.
\newblock A lower {W}egner estimate and bounds on the spectral shift function
  for continuum random {S}chr\"{o}dinger operators.
\newblock {\em J. Funct. Anal.}, 277(11):108284, 2019.

\bibitem[GHK07]{GerminetHK-07}
F.~Germinet, P.~D. Hislop, and A.~Klein.
\newblock Localization for {S}chr\"{o}dinger operators with {P}oisson random
  potential.
\newblock {\em J. Eur. Math. Soc. (JEMS)}, 9(3):577--607, 2007.

\bibitem[GK01]{GerminetK-01}
F.~Germinet and A.~Klein.
\newblock Bootstrap multiscale analysis and localization in random media.
\newblock {\em Comm. Math. Phys.}, 222(2):415--448, 2001.

\bibitem[GK04]{GerminetK-04}
F.~Germinet and A.~Klein.
\newblock A characterization of the {A}nderson metal-insulator transport
  transition.
\newblock {\em Duke Math. J.}, 124(2):309--350, 2004.

\bibitem[GK13]{GerminetK-13}
F.~Germinet and A.~Klein.
\newblock A comprehensive proof of localization for continuous {A}nderson
  models with singular random potentials.
\newblock {\em J. Eur. Math. Soc. (JEMS)}, 15(1):53--143, 2013.

\bibitem[GMP77]{GoldsheidMP-77}
I.~Ja. Gold\v{s}e{\u\i}d, S.~A. Mol\v{c}anov, and L.~A. Pastur.
\newblock A random homogeneous {S}chr\"odinger operator has a pure point
  spectrum.
\newblock {\em Funkcional. Anal. i Prilo\v zen.}, 11(1):1--10, 96, 1977.

\bibitem[GMRM15]{GerminetMRM-15}
F.~Germinet, P.~M{\"u}ller, and C.~Rojas-Molina.
\newblock Ergodicity and dynamical localization for {D}elone-{A}nderson
  operators.
\newblock {\em Rev. Math. Phys.}, 27(9):1550020, 36, 2015.

\bibitem[HM84]{HoldenM-84}
H.~Holden and F.~Martinelli.
\newblock On absence of diffusion near the bottom of the spectrum for a random
  {S}chr\"{o}dinger operator on {$L^{2}({\bf R}^{\nu })$}.
\newblock {\em Comm. Math. Phys.}, 93(2):197--217, 1984.

\bibitem[Kat76]{Kato-76}
T.~Kato.
\newblock {\em Perturbation {T}heory for {L}inear {O}perators}, volume 132 of
  {\em Grundlehren der Mathematischen Wissenschaften}.
\newblock Springer, Berlin, 2nd edition, 1976.

\bibitem[Kle08]{Klein-08}
A.~Klein.
\newblock Multiscale analysis and localization of random operators.
\newblock In {\em Random {S}chr\"o\-dinger {O}perators}, volume~25 of {\em
  Panor. Synth\`eses}, pages 121--159. Soc. Math. France, Paris, 2008.

\bibitem[Kle13]{Klein-13}
A.~Klein.
\newblock Unique continuation principle for spectral projections of
  {S}chr{\"o}dinger operators and optimal {W}egner estimates for non-ergodic
  random {S}chr{\"o}dinger operators.
\newblock {\em Comm. Math. Phys.}, 323(3):1229--1246, 2013.

\bibitem[KLNS12]{KloppLNS-12}
F.~Klopp, M.~Loss, S.~Nakamura, and G.~Stolz.
\newblock Localization for the random displacement model.
\newblock {\em Duke Math. J.}, 161(4):587--621, 2012.

\bibitem[Klo99]{Klopp-99}
F.~Klopp.
\newblock Internal {L}ifshits tails for random perturbations of periodic
  {S}chr\"{o}dinger operators.
\newblock {\em Duke Math. J.}, 98(2):335--396, 1999.

\bibitem[Klo02]{Klopp-02}
F.~Klopp.
\newblock Internal {L}ifshitz tails for {S}chr\"{o}dinger operators with random
  potentials.
\newblock {\em J. Math. Phys.}, 43(6):2948--2958, 2002.

\bibitem[KR00]{KloppR-00}
F.~Klopp and J.~Ralston.
\newblock Endpoints of the spectrum of periodic operators are generically
  simple.
\newblock {\em Methods Appl. Anal.}, 7(3):459--463, 2000.
\newblock Cathleen Morawetz: a great mathematician.

\bibitem[KSS98a]{KirschSS-98a}
W.~Kirsch, P.~Stollmann, and G.~Stolz.
\newblock Anderson localization for random {S}chr\"{o}dinger operators with
  long range interactions.
\newblock {\em Comm. Math. Phys.}, 195(3):495--507, 1998.

\bibitem[KSS98b]{KirschSS-98b}
W.~Kirsch, P.~Stollmann, and G.~Stolz.
\newblock Localization for random perturbations of periodic {S}chr\"odinger
  operators.
\newblock {\em Random Oper. Stoch. Equ}, 6:241--268, 1998.

\bibitem[KT16a]{KleinT-16a}
A.~Klein and C.~S.~S. Tsang.
\newblock Local behavior of solutions of the stationary {S}chr\"{o}dinger
  equation with singular potentials and bounds on the density of states of
  {S}chr\"{o}dinger operators.
\newblock {\em Comm. Partial Differential Equations}, 41(7):1040--1055, 2016.

\bibitem[KT16b]{KleinT-16}
A.~Klein and C.~S.~S. Tsang.
\newblock Quantitative unique continuation principle for {S}chr\"{o}dinger
  operators with singular potentials.
\newblock {\em Proc. Amer. Math. Soc.}, 144(2):665--679, 2016.

\bibitem[Kuc16]{Kuchment-16}
P.~Kuchment.
\newblock An overview of periodic elliptic operators.
\newblock {\em Bull. Amer. Math. Soc. (N.S.)}, 53(3):343--414, 2016.

\bibitem[KW02]{KloppW-02}
F.~Klopp and T.~Wolff.
\newblock Lifshitz tails for 2-dimensional random {S}chr\"{o}dinger operators.
\newblock {\em J. Anal. Math.}, 88:63--147, 2002.
\newblock Dedicated to the memory of Tom Wolff.

\bibitem[LPTV15]{LeonhardtPTV-15}
K.~Leonhardt, N.~Peyerimhoff, M.~Tautenhahn, and I.~Veseli\'{c}.
\newblock Wegner estimate and localization for alloy-type models with
  sign-changing exponentially decaying single-site potentials.
\newblock {\em Rev. Math. Phys.}, 27(4):1550007, 45, 2015.

\bibitem[MRM20]{RojasMolinaM-20}
P.~M{\"u}ller and C.~Rojas-Molina.
\newblock Localisation for delone operators via {B}ernoulli randomisation.
\newblock arXiv:2003.06325 [math-ph], 2020.

\bibitem[NTTV]{NakicSTTV-19}
I.~Naki\'c, M.~T\"aufer, M.~Tautenhahn, and I.~Veseli\'c.
\newblock Unique continuation and lifting of spectral band edges of
  {S}chr\"odinger operators on unbounded domains.
\newblock {\em To appear in J. Spectr. Theory}.
\newblock With an appendix by Albrecht Seelmann, arXiv:1804.07816 [math.SP],
  2018.

\bibitem[NTTV18]{NakicTTV-18}
I.~Naki\'c, M.~T\"aufer, M.~Tautenhahn, and I.~Veseli\'c.
\newblock Scale-free unique continuation principle, eigenvalue lifting and
  {W}egner estimates for random {S}chr\"odinger operators.
\newblock {\em Anal. PDE}, 11(4):1049--1081, 2018.

\bibitem[PF92]{PasturF-92}
L.~Pastur and A.~Figotin.
\newblock {\em Spectra of {Ra}ndom and {A}lmost-{P}eriodic {O}perators}, volume
  297 of {\em Grundlehren der Mathematischen Wissenschaften [Fundamental
  Principles of Mathematical Sciences]}.
\newblock Springer-Verlag, Berlin, 1992.

\bibitem[RM12a]{RojasMolina-12}
C.~Rojas-Molina.
\newblock Characterization of the {A}nderson metal-insulator transition for non
  ergodic operators and application.
\newblock {\em Ann. Henri Poincar\'e}, 13(7):1575--1611, 2012.

\bibitem[RM12b]{RojasMolina-Thesis}
C.~Rojas-Molina.
\newblock {\em Etude math\'ematique des propri\'et\'es de transport des
  op\'erateurs de Schr\"odinger al\'eatoires avec structure quasi-cristalline}.
\newblock PhD thesis, Universit\'e de Cergy-Pontoise, 2012.

\bibitem[RMV13]{RojasMolinaV-13}
C.~Rojas-Molina and I.~Veseli\'{c}.
\newblock Scale-free unique continuation estimates and applications to random
  {S}chr\"{o}dinger operators.
\newblock {\em Comm. Math. Phys.}, 320(1):245--274, 2013.

\bibitem[Sto01]{Stollmann-01}
P.~Stollmann.
\newblock {\em Caught by Disorder. Bound States in Random Media}.
\newblock Birkh\"auser Boston, Inc., Boston, MA, 2001.

\bibitem[SV17]{SchumacherV-17}
C.~Schumacher and I.~Veseli\'{c}.
\newblock Lifshitz tails for {S}chr\"{o}dinger operators with random breather
  potential.
\newblock {\em C. R. Math. Acad. Sci. Paris}, 355(12):1307--1310, 2017.

\bibitem[T{\"a}u18]{Taeufer-PhD}
M.~T{\"a}ufer.
\newblock {\em Quantitative unique continuation and applications}.
\newblock PhD thesis, Technische Universit\"at Dortmund, 2018.

\bibitem[TT18]{TaeuferT-18}
M.~T\"{a}ufer and M.~Tautenhahn.
\newblock Wegner estimate and disorder dependence for alloy-type {H}amiltonians
  with bounded magnetic potential.
\newblock {\em Ann. Henri Poincar\'{e}}, 19(4):1151--1165, 2018.

\bibitem[TV15]{TaeuferV-15}
M.~T\"{a}ufer and I.~Veseli\'{c}.
\newblock Conditional {W}egner estimate for the standard random breather
  potential.
\newblock {\em J. Stat. Phys.}, 161(4):902--914, 2015.

\bibitem[Ves02]{Veselic-02}
I.~Veseli\'{c}.
\newblock Localization for random perturbations of periodic {S}chr\"{o}dinger
  operators with regular {F}loquet eigenvalues.
\newblock {\em Ann. Henri Poincar\'{e}}, 3(2):389--409, 2002.

\bibitem[Ves08]{Veselic-08}
I.~Veseli\'c.
\newblock {\em Existence and {R}egularity {P}roperties of the {I}ntegrated
  {D}ensity of {S}tates of {R}andom {S}chr{\"o}dinger {O}perators}, volume 1917
  of {\em Lecture Notes in Mathematics}.
\newblock Springer-Verlag, Berlin, 2008.

\bibitem[Zen02]{Zenk-02}
Heribert Zenk.
\newblock Anderson localization for a multidimensional model including long
  range potentials and displacements.
\newblock {\em Rev. Math. Phys.}, 14(3):273--302, 2002.

\end{thebibliography}
\end{document}